\documentclass[final,5p,times,twocolumn]{elsarticle}

\usepackage[utf8]{inputenc}        
\usepackage[english]{babel} 
\usepackage{amssymb}
\usepackage{amsmath}
\makeatother
\usepackage{color}
\usepackage[usenames,dvipsnames,svgnames,table]{xcolor}
\usepackage{colortbl}
\colorlet{RED}{red}
\usepackage{graphicx,subfigure}
\graphicspath{{./}}
\usepackage{epstopdf}
\usepackage{wrapfig} 
\makeatletter
\usepackage{babelbib}
\biboptions{compress}
\begin{document}
 \title{Plasma electron trapping in quasistatic simulations of plasma wakefield acceleration}
 \author[BINP,NSU]{P.V. Tuev}
 \author[BINP,NSU]{A.P. Sosedkin}
 \author[BINP,NSU]{K.V. Lotov}
 \address[BINP]{Budker Institute of Nuclear Physics SB RAS, 630090 Novosibirsk, Russia}
 \address[NSU]{Novosibirsk State University, 630090 Novosibirsk, Russia}
 \begin{abstract}
     Plasma wakefield acceleration studies currently rely considerably 
     on  simulating this effect numerically using highly specialized software. 
     Exorbitant computational difficulty of the problem requires simplifying models 
     and methods, limiting such software applicability. Quasistatic approximation, 
     for example, utilizes a plasma model that does not include trapping plasma 
     electrons by the wakefield. This article presents a method that reuses 
     a quasistatic plasma-beam solver to calculate parameters of wakefield-trapped plasma electrons.

 \end{abstract}
 \maketitle

\section{Introduction}

Novel acceleration methods are a hot topic of research in particle acceleration,
currently spearheaded by plasma wakefield acceleration techniques \cite{RAST9-19,RAST9-63,RAST9-85}. 
The amplitude of the electric field in plasma waves exceeds those attainable in conventional radiofrequency structures by several orders of magnitude. 
This could reduce the size of the acceleration section and, as result, the total cost of the system. 
Experimental research pursues two goals: developing future colliders for high energy physics \cite{RAST9-209}
and designing compact electron and ray sources for scientific and commercial applications \cite{PPCF56-084015,NIMA-829-291,NatPhys4-447}.

Laser wakefield acceleration is more suitable for the latter goal.
Modern laser systems are able to produce laser pulses capable of driving non-linear plasma waves. 
Such scheme allows to forego external injection and accelerate trapped plasma electrons \cite{APB74-355,RMP81-1229}.

Diagnosing of the wakefield formation is a complicated task
due to the micron scale of plasma wave structures.
Thus, numerical simulations are crucial for understanding this phenomenon.
Such simulations are computationally demanding, thus
development of efficient approaches, models and algorithms is necessary.
One of these models, quasistatic approximation \cite{PoP4-217}, significantly reduces the requirements for
computing resources in comparison with full particle in cell (PIC) simulations.
Unfortunately, the plasma model in this approximation does not take plasma electron trapping into account.  

In this paper we describe a method of accounting for electron trapping in the quasistatic approximation. 
We present a quick review of quasistatic simulation principles in Section \ref{quasist}.
Then we discuss an algorithm for trapping simulation is discussed in Section \ref{trapping_in_lcode}
and benchmark the suggested method against PIC simulation in Section \ref{comp_osiris}.

\section{Description of quasistatic approximation}\label{quasist}
Our algorithm has been developed using 2D code LCODE \cite{PoP5-785, PRST-AB6-061301, NIMA-829-350}, but it should be applicable to other quasistatic codes as well.
We use comoving axisymmetric coordinates $(r, \xi)$, where $r$ is the transverse coordinate,
$\xi = z - ct$, $z$ is the axis of driver propagation, $c$ is the speed of light, and
$t$ is time in the laboratory frame of reference.

\begin{figure}[tbh!]
    \center{\includegraphics[width=0.95\linewidth]{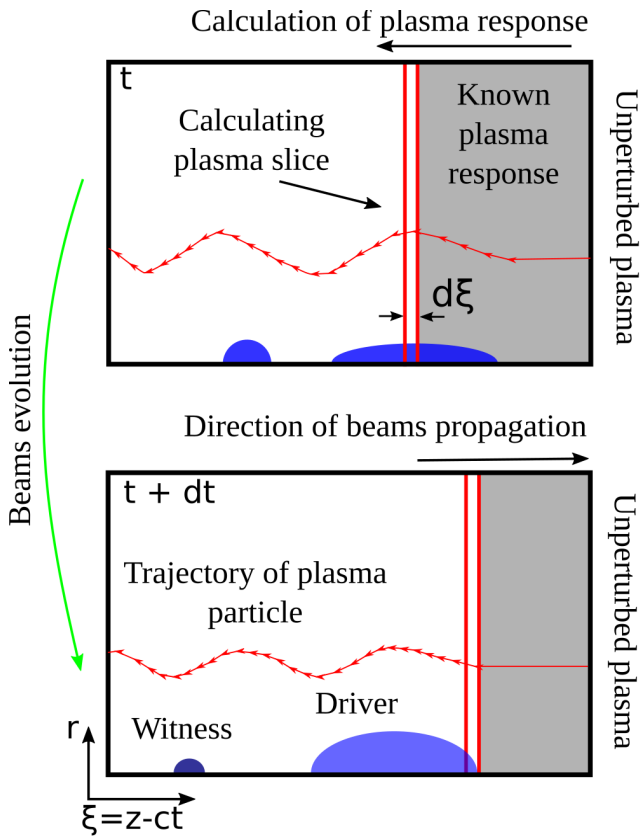}}
    \caption{A schematic description of a simulation in the quasistatic approximation.}
    \label{lcode_scheme}
\end{figure}

The scheme of the simulation is shown in Fig.~\ref{lcode_scheme}.
Simulation window moves with the speed of light in the direction of driver propagation, thus a time step $dt$ corresponds to the space step $dz = c~dt$.

The driver evolves slowly, and
perturbations of adjacent plasma slices are nearly identical.
Because of this, it is sufficient to calculate the response of a single layer as the driver passes through it,
and then proceed to the next layer only as the driver propagates further into the plasma and evolves significantly.
For any given propagation distance $ct$ the plasma response depends on $\xi$ - 
position behind the driver front and can be calculated slice by slice against this axis.
The equations of plasma particle motion are
\begin{equation}\label{pp_motion}
\begin{gathered}
    \frac{d\vec{p}_{p}}{d\xi} = \frac{q_{p}}{v_{pz} - c} \left( \vec{E} +\frac{1}{c} \left[\vec{v}_{p} \times \vec{B}\right] \right), \\
    \frac{dr_{p}}{d\xi} = \frac{v_{pr}}{v_{pz} - c}, \qquad 
    \vec{v}_{p} = \frac{c\vec{p}_{p}}{\sqrt{M^2_{p}c^2 + p^2_{p}}},
\end{gathered}
\end{equation}
where $\vec{p}_{p}$ and $\vec{v}_{p}$ are momentum and velocity, $M_{p}$ and $q_{p}$ are mass and charge, 
$r_{p}$ is the radial position of plasma particles. 
Plasma macroparticles describe a group of real particles with the same initial transverse coordinate and momentum 
and located between $z$ and $z+dz$. 
The plasma affects the driver and the witness for a time step $dt$, which is much larger than step $d\xi/c$ used for simulating plasma response.
The next time step is repeated with evolved driver passing through another layer of initially unperturbed plasma.  
Quasistatic approximation allows to speed up the simulation in comparison to generic PIC codes 
by the ratio of driver time step $dt$ to the spatial resolution $d\xi/c$.

Charged beams are simulated with fully relativistic macroparticles. 
The equations of beam particle motion in the comoving window are
\begin{equation}\label{b_motion}
\begin{gathered}
    \vec{v}_{b} = \frac{c\vec{p}_{b}}{\sqrt{m^2_{b}c^2 + p^2_{b}}}, \qquad
    \frac{d\xi_{b}}{dt} = v_{bz} - c, \\ 
    \frac{dr_{b}}{dt} = v_{br}, \qquad 
    \frac{d\vec{p}_{b}}{dt} = q_{b} \left( \vec{E} +\frac{1}{c} \left[\vec{v}_{b} \times \vec{B}\right] \right), 
\end{gathered}
\end{equation}
where $\vec{p}_{b}$ and $\vec{v}_{b}$ are momentum and velocity, 
$M_{b}$ and $q_{b}$ are mass and charge, $r_{b}$ and $\xi_b$ are coordinates of beam particles in the simulation window \cite{PRST-AB6-061301}. 

In our simulations the wakefield is driven by a laser pulse.
Laser evolution is described by the laser vector-potential envelope $A(r,\xi,t)$. 
The envelope evolves according to the Maxwell equations in quasistatic approximation \cite{PoP4-217}.
Terms of the order of  $(k_p/k_0)^2$ have been neglected, where $k_0$ is the laser wave number, 
$k_p = \omega_p/c$, $\omega_p = \sqrt{4\pi n_0 e^2/m_e}$ is the plasma frequency, $n_0$ is the unperturbed plasma density,
$e$ and $m_e$ are electron charge and mass. 
The laser pulse affects only plasma electrons through the ponderomotive force.  
The initial laser shape~is 
\begin{equation}\label{laser_def}
\begin{gathered}
    A^2 = A_0 e^{-2r^2/r_0^2}
    \begin{cases}
        1 + \cos \left( \frac{\sqrt{\pi}(\xi-\xi_c)}{c\tau} \right), & |\xi - \xi_c| < \sqrt{\pi}c\tau, \\
        0, & \text{otherwise}, 
    \end{cases} \\
    A_0 = \frac{m_e^2c^4}{e^2} \frac{a_0^2}{2},
\end{gathered}
\end{equation}
where $a_0$ is the maximum of $A$ in units of $m_e^2c^4/e^2$,  $r_0$ and $\tau$ are pulse spot size and duration, 
$\xi_c$ is the position of pulse center in the simulation window. 

\section{Algorithm for simulating plasma electron trapping} \label{trapping_in_lcode}

The quasistatic approach does not take several physical phenomena into account.
In particular, the plasma electron trapping is not described by Eq.~\eqref{pp_motion}.
Plasma electron trapping occurs when the accelerating fields in plasma wave reach several units of the cold wavebreaking field $E_0=m_e\omega_p c/e$. 
We assume that trapped electrons do not alter the wakefield structure much because they only constitute a small part of all plasma electrons \cite{PoP13-033103}.
The state of all plasma particles is calculated for each $\xi$-slice in quasistatic approach. 
Let us apply the beam model [Eq.~\eqref{b_motion}] to simulating plasma electrons.
To this end, we replicate them as beam test particles, which do not change the wakefield, at the distance $\xi$ behind the laser pulse.
\begin{figure}[h!]
    \center{\includegraphics[width=1\linewidth]{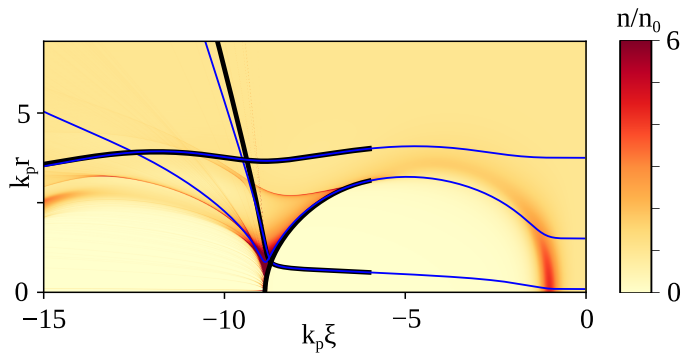}}
    \caption{Trajectories of plasma electrons (thin blue lines) and test beam particles (thick black lines). 
            Background colors show the plasma electron density $n$.} 
    \label{traj}
\end{figure}

Typical trajectories of plasma electrons and corresponding beam particles are shown in Fig.~\ref{traj}.
A laser pulse with $a_0 = 3.5$, $\tau = \omega_p^{-1}$, $r_0 = 2.6k_p^{-1}, k_0 = 30k_p$ is used as a driver.
The energy gain of some electrons is significant, and the quasistatic approximation is no longer applicable for them.  
They stay in the simulation window according to complete equations of motion \eqref{b_motion} and form a witness.   
Nevertheless, the solutions for equations \eqref{b_motion} and \eqref{pp_motion} are equivalent for the most of the electrons.

Simulation of witness formation consists of two steps. 
First, the plasma state is calculated using the quasistatic approximation.
After that, the test beam particles are added 
to the witness beam formed at previous time steps.  
Note that a test particle is generated for each plasma electron involved in wake formation. 
Implementing these two steps allows to simulate witness formation and acceleration along the driver propagation.

This algorithm if valid if the witness does not affect the plasma wake. 
The applicability of the method can be extended 
to longer acceleration distances by taking the witness charge into account. 
However, direct conversion of a plasma particle into a beam particle must be accompanied 
by abrupt termination of the plasma particle's trajectory in the simulation window. 
Multiple events of this kind makes  the plasma solver unstable.
Thus, the correct conversion from
one sort of particles into another requires additional study and is not discussed in this work. 

\section{Comparison with PIC}\label{comp_osiris}

A comparison with known results is necessary for verifying such algorithms.
We have compared our simulation with Ref.~\cite{PRST-AB10-061301},
which simulates the interaction of a high intensity laser pulse with an uniform plasma using PIC code OSIRIS \cite{Fonseca2002}. 
Laser pulse with $a_0 = 4$ ensures trapping from the beginning of the plasma. 
The duration and spot size of the pulse are $\tau = 17$~fs and $r_0 = 20~\mu$m. 
The simulated plasma has a length of 7.5~mm and density $n_0 = 1.5 \times 10^{18} \text{cm}^{-3}$.

\begin{figure}[h]
    \begin{minipage}[h]{0.49\linewidth}   
        \includegraphics[width=1\linewidth]{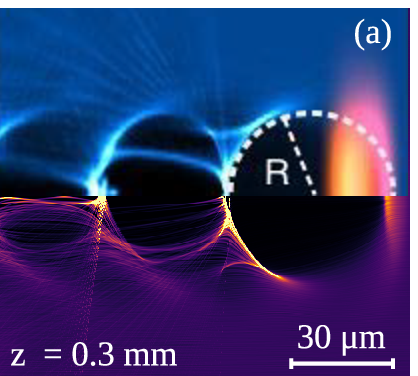} 
    \end{minipage}
    \hfill
    \begin{minipage}[h]{0.49\linewidth}   
        \includegraphics[width=1\linewidth]{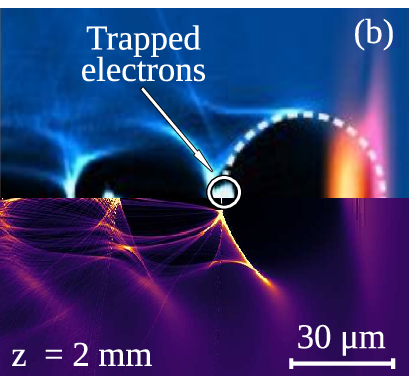} 
    \end{minipage}
    \caption{ The electron plasma density for OSIRIS (top) \cite{PRST-AB10-061301} and LCODE (bottom) simulations after
              0.3~mm~(a) and  2~mm~(b).
             Top pictures also show the laser pulse.}
    \label{dens_vs_osiris}
\end{figure}

The structure of the resulting plasma wave 
is similar in both simulations  
until the influence the witness charge becomes significant and 
changes the wakefield structure (Fig.~\ref{dens_vs_osiris}).
In the region ahead of the the witness, the quasistatic simulation 
always matches PIC results. 

\begin{figure}[h!]
    \begin{minipage}[h]{0.476\linewidth}   
        \includegraphics[width=1\linewidth]{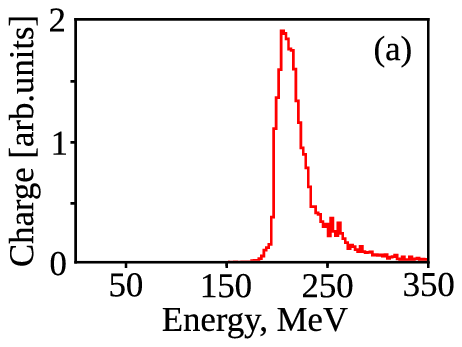} 
    \end{minipage}   
    \hfill
    \begin{minipage}[h]{0.424\linewidth}
        \includegraphics[width=1\linewidth]{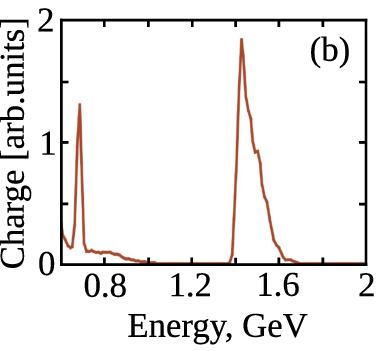}
    \end{minipage}    
    \caption{Energy spectrum of the trapped beams: (a) LCODE simulation after 1~mm of plasma,
          (b) OSIRIS simulation after 7.5~mm of plasma \cite{PRST-AB10-061301}}
    \label{spectr_comp}
\end{figure}

We also compare the energy spectra of the trapped electrons (Fig.~\ref{spectr_comp}).
Since the first bubble is least influenced by trapped electrons, let us consider electron trapping only in this wake period.  
We also reduce the interaction distance to 1~mm, in comparison with 7.5~mm in Ref.~\cite{PRST-AB10-061301},
to avoid the influence of the witness space charge on the wakefield.
Despite the fact that we have neglected the fields of trapped plasma electrons, 
we take into account their relative charges in order to obtain the correct witness energy spectrum. 
We compare this spectrum with the main peak in Fig.~\ref{spectr_comp}(b) that corresponds to electrons trapped at the first wake period. 

The average witness energy grows to 224 MeV after 1~mm interaction length simulated with the quasistatic code. 
Scaling this value up to 7.5 mm interaction length yields a result which exceeds the OSIRIS result (1.5 GeV) by 10\%. 
This mismatch probably comes from decrease of the accelerating field at the end of interaction,
caused by laser pulse depletion. 
The shapes of both spectra are quite similar: both feature a sharp edge at the lower energy and a high-energy long tail.

\section{Conclusion}

The possibility of electron trapping simulation with a quasistatic code has been demonstrated. 
The quasistatic approximation is valid for most of the plasma electrons.
Plasma wake can be calculated with a quasistatic code 
as long as the contribution of trapped electrons to the charge density is negligible. 
The scaling of simulation results matches to PIC simulation made with OSIRIS. 
The algorithm suggested in this paper may be
applicable to other quasistatic codes.

\section*{Acknowledgments}
The reported study was funded by RFBR and Government of the Novosibirsk region according to the research project No. 17-41-543162.

\section*{References}
\bibliography{nima_2017}
\end{document}